\documentstyle[12pt]{article}

\catcode`\@=11
\long\def\@makefntext#1{ %\parindent 1em
\protect\noindent \hbox to 3.2pt {\hskip-.9pt
$^{{\ninerm\@thefnmark}}$\hfil}#1\hfill} %can be used

\def\thefootnote{\fnsymbol{footnote}}
 \def\@makefnmark{\hbox to 0pt{$^{\@thefnmark}$\hss}}  %original

\def\ps@myheadings{\let\@mkboth\@gobbletwo
\def\@oddhead{\hbox{} %\sl
\rightmark\hfil\ninerm\thepage}
\def\@oddfoot{}\def\@evenhead{\ninerm\thepage\hfil %\sl
\leftmark\hbox{}}\def\@evenfoot{}
\def\sectionmark##1{}\def\subsectionmark##1{}}

\newcommand{\beq}{\begin{equation}}
\newcommand{\eeq}{\end{equation}}
\newcommand{\bea}{\begin{eqnarray}}
\newcommand{\eea}{\end{eqnarray}}
\newcommand{\barr}{\begin{array}}
\newcommand{\earr}{\end{array}}
\newcommand{\bc}{\begin{center}}
\newcommand{\ec}{\end{center}}
\newcommand{\btab}{\begin{tabular}}
\newcommand{\etab}{\end{tabular}}
\newcommand{\gv}{\mbox{GeV}}

\newcommand{\nn}{\nonumber}
\newcommand{\ra}{\rightarrow}

\newcommand{\sz}{\Sigma^{ZZ}}
\newcommand{\sw}{\Sigma^{WW}}

\newcommand{\dro}{\Delta\rho}
\newcommand{\drqcd}{\delta\!\rho\,_{QCD}}
\newcommand{\roro}{\rho^{(2)}}

\newcommand{\drb}{\Delta\overline{\rho}}

\newcommand{\al}{\alpha}

\newcommand{\g}{\gamma}
\newcommand{\G}{\Gamma}
\newcommand{\Gmu}{G_{\mu}}

\newcommand{\ganu}{\gamma_{\nu}}

\newcommand{\gafi}{\gamma_5}

\newcommand{\Pig}{\Pi^{\gamma}}

\newcommand{\noi}{\noindent}

\newcommand{\sm}{Standard Model }
\newcommand{\su}{SU(2)$\times$U(1) }

\newcommand{\dal}{\Delta\alpha}

\newcommand{\mz}{M_Z^2}
\newcommand{\mw}{M_W^2}

\newcommand{\real}{\mbox{Re}}

\newcommand{\Dr}{\Delta r}

\newcommand{\alr}{A_{LR}}

\newcommand{\ass}{asymmetries }

\newcommand{\pr}{Phys.\ Rev.\ }
 \newcommand{\prd}{Phys.\ Rev.\ D }
\newcommand{\zp}{Z.\ Phys.\ C }
\newcommand{\plb}{Phys.\ Lett.\ B }
 \newcommand{\prl}{Phys.\ Rev.\ Lett.\ }
\newcommand{\np}{Nucl.\ Phys.\ B }

\newcommand{\ms}{\overline{MS}}

\begin{document}

%----------------------------PROCSLA.STY---------------------------------------
\newcommand{\symbolfootnote}{\renewcommand{\thefootnote}
        {\fnsymbol{footnote}}}
\renewcommand{\thefootnote}{\fnsymbol{footnote}}
\newcommand{\alphfootnote}
        {\setcounter{footnote}{0}
         \renewcommand{\thefootnote}{\sevenrm\alph{footnote}}}

%------------------------------------------------------------------------------
%NEW DEFINED SECTION COMMANDS
\newcounter{sectionc}\newcounter{subsectionc}\newcounter{subsubsectionc}
\renewcommand{\section}[1] {\vspace{0.6cm}\addtocounter{sectionc}{1}
\setcounter{subsectionc}{0}\setcounter{subsubsectionc}{0}\noindent
        {\bf\thesectionc. #1}\par\vspace{0.4cm}}
\renewcommand{\subsection}[1] {\vspace{0.6cm}\addtocounter{subsectionc}{1}
        \setcounter{subsubsectionc}{0}\noindent
        {\it\thesectionc.\thesubsectionc. #1}\par\vspace{0.4cm}}
\renewcommand{\subsubsection}[1]
{\vspace{0.6cm}\addtocounter{subsubsectionc}{1}
        \noindent {\rm\thesectionc.\thesubsectionc.\thesubsubsectionc.
        #1}\par\vspace{0.4cm}}
\newcommand{\nonumsection}[1] {\vspace{0.6cm}\noindent{\bf #1}
        \par\vspace{0.4cm}}

%NEW MACRO TO HANDLE APPENDICES
\newcounter{appendixc}
\newcounter{subappendixc}[appendixc]
\newcounter{subsubappendixc}[subappendixc]
\renewcommand{\thesubappendixc}{\Alph{appendixc}.\arabic{subappendixc}}
\renewcommand{\thesubsubappendixc}
        {\Alph{appendixc}.\arabic{subappendixc}.\arabic{subsubappendixc}}

\renewcommand{\appendix}[1] {\vspace{0.6cm}
        \refstepcounter{appendixc}
        \setcounter{figure}{0}
        \setcounter{table}{0}
        \setcounter{equation}{0}
        \renewcommand{\thefigure}{\Alph{appendixc}.\arabic{figure}}
        \renewcommand{\thetable}{\Alph{appendixc}.\arabic{table}}
        \renewcommand{\theappendixc}{\Alph{appendixc}}
        \renewcommand{\theequation}{\Alph{appendixc}.\arabic{equation}}
%       \noindent{\bf Appendix \theappendixc. #1}\par\vspace{0.4cm}}
        \noindent{\bf Appendix \theappendixc #1}\par\vspace{0.4cm}}
\newcommand{\subappendix}[1] {\vspace{0.6cm}
        \refstepcounter{subappendixc}
        \noindent{\bf Appendix \thesubappendixc. #1}\par\vspace{0.4cm}}
\newcommand{\subsubappendix}[1] {\vspace{0.6cm}
        \refstepcounter{subsubappendixc}
        \noindent{\it Appendix \thesubsubappendixc. #1}
        \par\vspace{0.4cm}}

%------------------------------------------------------------------------------
%MARCO FOR ABSTRACT BLOCK
\def\abstracts#1{{
        \centering{\begin{minipage}{30pc}\tenrm\baselineskip=12pt\noindent
        \centerline{\tenrm ABSTRACT}\vspace{0.3cm}
        \parindent=0pt #1
        \end{minipage} }\par}}

%------------------------------------------------------------------------------
%NEW MACRO FOR BIBLIOGRAPHY
\newcommand{\bibit}{\it}
\newcommand{\bibbf}{\bf}
\renewenvironment{thebibliography}[1]
        {\begin{list}{\arabic{enumi}.}
        {\usecounter{enumi}\setlength{\parsep}{0pt}
%1.25cm IS STRICTLY FOR PROCSLA.TEX ONLY
\setlength{\leftmargin 1.25cm}{\rightmargin 0pt}
%0.52cm IS FOR NEW DATA FILES
%\setlength{\leftmargin 0.52cm}{\rightmargin 0pt}
         \setlength{\itemsep}{0pt} \settowidth
        {\labelwidth}{#1.}\sloppy}}{\end{list}}

%------------------------------------------------------------------------------
%FOLLOWING THREE COMMANDS ARE FOR 'LIST' COMMAND.
\topsep=0in\parsep=0in\itemsep=0in
\parindent=1.5pc

%LIST ENVIRONMENTS
\newcounter{itemlistc}
\newcounter{romanlistc}
\newcounter{alphlistc}
\newcounter{arabiclistc}
\newenvironment{itemlist}
        {\setcounter{itemlistc}{0}
         \begin{list}{$\bullet$}
        {\usecounter{itemlistc}
         \setlength{\parsep}{0pt}
         \setlength{\itemsep}{0pt}}}{\end{list}}

\newenvironment{romanlist}
        {\setcounter{romanlistc}{0}
         \begin{list}{$($\roman{romanlistc}$)$}
        {\usecounter{romanlistc}
         \setlength{\parsep}{0pt}
         \setlength{\itemsep}{0pt}}}{\end{list}}

\newenvironment{alphlist}
        {\setcounter{alphlistc}{0}
         \begin{list}{$($\alph{alphlistc}$)$}
        {\usecounter{alphlistc}
         \setlength{\parsep}{0pt}
         \setlength{\itemsep}{0pt}}}{\end{list}}

\newenvironment{arabiclist}
        {\setcounter{arabiclistc}{0}
         \begin{list}{\arabic{arabiclistc}}
        {\usecounter{arabiclistc}
         \setlength{\parsep}{0pt}
         \setlength{\itemsep}{0pt}}}{\end{list}}

%------------------------------------------------------------------------------
%FIGURE CAPTION
\newcommand{\fcaption}[1]{
        \refstepcounter{figure}
        \setbox\@tempboxa = \hbox{\tenrm Fig.~\thefigure. #1}
        \ifdim \wd\@tempboxa > 6in
           {\begin{center}
        \parbox{6in}{\tenrm\baselineskip=12pt Fig.~\thefigure. #1 }
            \end{center}}
        \else
             {\begin{center}
             {\tenrm Fig.~\thefigure. #1}
              \end{center}}
        \fi}

%TABLE CAPTION
\newcommand{\tcaption}[1]{
        \refstepcounter{table}
        \setbox\@tempboxa = \hbox{\tenrm Table~\thetable. #1}
        \ifdim \wd\@tempboxa > 6in
           {\begin{center}
        \parbox{6in}{\tenrm\baselineskip=12pt Table~\thetable. #1 }
            \end{center}}
        \else
             {\begin{center}
             {\tenrm Table~\thetable. #1}
              \end{center}}
        \fi}

%------------------------------------------------------------------------------
%ACKNOWLEDGEMENT: this portion is from John Hershberger
\def\@citex[#1]#2{\if@filesw\immediate\write\@auxout
        {\string\citation{#2}}\fi
\def\@citea{}\@cite{\@for\@citeb:=#2\do
        {\@citea\def\@citea{,}\@ifundefined
        {b@\@citeb}{{\bf ?}\@warning
        {Citation `\@citeb' on page \thepage \space undefined}}
        {\csname b@\@citeb\endcsname}}}{#1}}

\newif\if@cghi
\def\cite{\@cghitrue\@ifnextchar [{\@tempswatrue
        \@citex}{\@tempswafalse\@citex[]}}
\def\citelow{\@cghifalse\@ifnextchar [{\@tempswatrue
        \@citex}{\@tempswafalse\@citex[]}}
\def\@cite#1#2{{$\null^{#1}$\if@tempswa\typeout
        {IJCGA warning: optional citation argument
        ignored: `#2'} \fi}}
\newcommand{\citeup}{\cite}

%------------------------------------------------------------------------------
%FOR FNSYMBOL FOOTNOTE AND ALPH{FOOTNOTE}
\def\fnm#1{$^{\mbox{\scriptsize #1}}$}
\def\fnt#1#2{\footnotetext{\kern-.3em
        {$^{\mbox{\sevenrm #1}}$}{#2}}}

%------------------------------------------------------------------------------
\font\twelvebf=cmbx10 scaled\magstep 1
\font\twelverm=cmr10 scaled\magstep 1
\font\twelveit=cmti10 scaled\magstep 1
\font\elevenbfit=cmbxti10 scaled\magstephalf
\font\elevenbf=cmbx10 scaled\magstephalf
\font\elevenrm=cmr10 scaled\magstephalf
\font\elevenit=cmti10 scaled\magstephalf
\font\bfit=cmbxti10
\font\tenbf=cmbx10
\font\tenrm=cmr10
\font\tenit=cmti10
\font\ninebf=cmbx9
\font\ninerm=cmr9
\font\nineit=cmti9
\font\eightbf=cmbx8
\font\eightrm=cmr8
\font\eightit=cmti8

%----------------------START OF DATA FILE------------------------------
\vspace*{-1.5cm}
%\hspace*{\fill} KA-TP-2-1995
\hspace*{11.0cm} KA-TP-2-1995

\vspace*{1.5cm}
\centerline{\tenbf ELECTROWEAK PRECISION OBSERVABLES - }
\baselineskip=16pt
\centerline{\tenbf AN INDIRECT ACCESS TO THE TOP QUARK
             \footnote{talk at the 138.\ Heraeus Seminar on Heavy Quarks
              Physics, Bad Honnef, Germany, December 1994} }
%\baselineskip=22pt
%\centerline{\tenbf INSTRUCTIONS FOR TYPESETTING CAMERA-READY}
%\baselineskip=16pt
%\centerline{\tenbf MANUSCRIPT USING COMPUTER SOFTWARE}
%\centerline{\ninerm (For 20\% Reduction to 6 in. $\times$ 8.5 in. Trim Size)}
\vspace{0.8cm}
\centerline{\tenrm WOLFGANG HOLLIK}
                  \baselineskip=13pt
\centerline{\tenit Institut f\"ur Theoretische Physik,
                   Universit\"at Karlsruhe}
\baselineskip=12pt
\centerline{\tenit D-76128 Karlsruhe, Germany}
\vspace{0.9cm}
\abstracts{The role of the top quark in the Standard Model predictions
for  electroweak precision observables
is reviewed and the implications of the experimental data for the
indirect determination of the top
mass range is discussed.}

\vspace*{1cm}
%\vfil
%\vspace{0.8cm}
\twelverm   %modified by CLee 23/07/93
\baselineskip=14pt
\section{Electroweak precision observables}
%\vspace*{-0.7cm}
%\subsection{The fermion families}
%\vspace*{-0.35cm}
%\vglue 0.4cm
%\vglue 0.3cm
%\leftline{\twelveit 1.1. The fermion families}
%\vglue 0.4cm
%\vglue 1pt
The top quark completes the third generation of fermions and makes
the \sm anomaly free. The experimental observation of the top
has been announced recently by the CDF collaboration \cite{top}
with mass
 $m_t = 176 \pm 8\pm 10$ GeV and by the D0 collaboration \cite{d0} with
$m_t = 199^{+19}_{-21}\pm 22$ GeV.

On the other hand, indirect information about the top quark is obtained
from electroweak precision observables.
The possibility of performing precision tests of the electroweak theory
is based
on the formulation of the \sm as a renormalizable quantum field
theory preserving its predictive power beyond tree level
calculations. With the experimental accuracy in the investigation
of the fermion-gauge boson interactions being sensitive to the loop
induced quantum effects, especially the heavy fermion sector of
the \sm is probed.

Before one can make predictions from the theory,
a set of independent parameters has to be determined from experiment.
All the practical schemes make use of the same physical input quantities
$ \al, \; \Gmu,\; M_Z,\; m_f,\; M_H $
for fixing the free parameters of the SM.
In terms of these the set of precisely measureable quantities $M_W$ and
$ \Gamma_Z, \; \Gamma_f, A_{FB}, \; A_{LR}, \; A^{pol}_{\tau},
\cdots $ at the $Z$ resonance can be calculated as predictions depending
on $m_t$ and $M_H$, together with the strong coupling $\al_s$.

\medskip
%\subsection{The vector boson and Higgs sector}
%\vglue 0.4cm
%\vglue 0.3cm
%\leftline{\twelveit 1.2. The vector boson and Higgs sector}
%\vglue 0.4cm
%\vglue 1pt
The spectrum of the vector bosons $\g,W^{\pm},Z$  with
masses       \cite{LEP,pp}
 \beq
  M_W = 80.23\pm 0.18\, \gv, \;\;\;\;
  M_Z = 91.1888\pm 0.0044\, \gv
\eeq
is reconciled with the \su local gauge symmetry with the help of the
Higgs mechanism. For a general structure of the scalar sector,
the electroweak mixing angle is related to the vector boson
masses  by
\beq
 s^2_{\theta} \equiv \sin^2\theta =
 1-\frac{\mw}{\rho\mz} =
   1-\frac{\mw}{\mz} + \frac{\mw}{\mz} \dro
 \equiv s_W^2 + c_W^2 \dro
\eeq
where the $\rho$-parameter  $\rho = (1-\dro)^{-1}$
is an additional free parameter.
In models with scalar doublets only, in particular in the
minimal model, one has the tree level relation $\rho=1$.
Loop effects, however, induce a deviation $\dro \neq 0$.

\medskip
The Standard Model prediction for
$\dro$ from radiative corrections is dominated
by the $(t,b)$ doublet contribution \cite{rho},
 in 1-loop and
 neglecting $m_b$:
\beq
   \dro \, =  \,
     \frac{\sz(0)}{\mz} -\frac{\sw(0)}{\mw}
              \, \simeq \,
        \frac{3\Gmu m_t^2}{8\pi^2\sqrt{2}}
    \, \equiv \, 3 x_t \, .
\eeq
This  large
contribution constitutes  the leading
shift for the electroweak mixing angle
when inserted into Eq.\ (2).

\smallskip \noi
Another large loop effect in the electroweak parameters is due to
the fermionic content of the subtracted photon vacuum polarization
\cite{vacpol}
\beq
 \dal =   \Pig_{ferm}(0) -
     \real\,\Pig_{ferm}(\mz)
      =  0.0593\pm 0.0007
\eeq
as a recent update confirming essentially  the previous result
\cite{vacpol0}
but with a sightly smaller error.
It  corresponds to a QED induced shift
in the electromagnetic fine structure constant yielding
an effective fine structure constant at the $Z$ mass scale:
$$
   \al(\mz) \, =\, \frac{\al}{1-\dal}\,=\,
   \frac{1}{128.9\pm 0.1} \, .
$$

\medskip
\section{The vector boson masses}
%\vglue 0.4cm
%\vglue 0.3cm
%\leftline{\twelveit 2.1. The vector boson masses}
%\vglue 0.4cm
%
%\vglue 1pt
The correlation between
the masses $M_W,M_Z$ of the vector bosons          in terms
of the Fermi constant $\Gmu$       reads in 1-loop order
of the Standard Model \cite{sirmar}:
\beq
\frac{\Gmu}{\sqrt{2}}   =
            \frac{\pi\al}{2s_W^2 M_W^2} \left[
        1+ \Dr(\al,M_W,M_Z,M_H,m_t)\right]\, .
\eeq
The 1-loop correction $\Dr$ can be written in the following way
\beq
 \Dr = \Delta\al -\frac{c_W^2}{s_W^2}\,\dro
         + (\Dr)_{remainder} \, .
\eeq
in order to separate the
leading fermionic contributions
                $\dal$ and $\dro$.
All other terms are collected in
$(\Dr)_{remainder}$,
the typical size of which is of the order $\sim 0.01$.

\bigskip
The presence of large terms in $\Dr$ requires the consideration
of higher than 1-loop effects.
The modification of Eq.\ (5) according to
\beq
         1+\Dr  \, \ra\, \frac{1}{(1-\Delta\al)\cdot
(1+\frac{c_W^2}{s_W^2}\drb) \, -\,(\Dr)_{remainder}}
 \equiv \frac{1}{1-\Dr}
\eeq
with
\beq
    \drb  =   3\,x_t
         \cdot \left[1+
           x_t\,\roro)
 + \drqcd  \right]
\eeq
accommodates, besides the leading log resummation of $\dal$,
the resummation of the leading top contribution \cite{chj}
 in terms
of $\drb$ which contains also irreducible higher order parts:
the electroweak 2-loop contribution
$\roro(M_H/m_t)$  \cite{barbieri}, and the
QCD correction $\drqcd$ up to $O(\al\al_s^2)$ \cite{djouadi,tarasov}.
The complete
 $O(\al\al_s)$ corrections to the self energies
 beyond the $m_t^2$ approximation are available from
 perturbative  calculations
\cite{qcd} and by means of dispersion relations \cite{dispersion1}
(see also \cite{kniehl93}).

\smallskip \noi
 The quantity $\Dr$ in Eq.\ (7)
$$
\Dr \,=\, 1\,-\,\frac{\pi\al}{\sqrt{2}\Gmu} \, \frac{1}
     {M_W^2 \left(1-\frac{M_W^2}{M_Z^2} \right) } \, .
$$   is
experimentally  determined by $M_Z$ and
$M_W$.
Theoretically, it is computed from $M_Z,\Gmu,\al$
after specifying the masses $M_H,m_t$.
The theoretical prediction for $\Dr$
is displayed in Figure 1.
For comparison with data, the experimental $1\sigma$ limits
from the direct measurements of $M_Z$ at LEP and
$M_W$ in $p\bar{p}$ are indicated, fully consistent with the recent
direct top mass determination and with the standard theory.

\begin{figure} % fig 2
\vspace*{3.25in}
% next line was used to print actual photo, commented out here.
% your syntax will probably differ.
% \hbox to\hsize{\hfill\special{ps: epsfile photo.eps}\kern3in\hfill}
\caption{ $\Dr$ as a function of the top mass for
          $M_H=60$ and  $1000$ GeV.
          $1\sigma$ bounds from $M_Z$ and $s_W^2$:
          horizontal band from $p\bar{p}$, $\bullet$ from $\nu N$.}
\end{figure}

The quantity $s_W^2$ resp.\  the ratio $M_W/M_Z$
can indirectly be measured in deep-inelastic neutrino scattering,
yielding the present
world average \cite{bodek}
 consistent with the direct vector boson mass measurements:
 $$ s_W^2 =  0.2256 \pm 0.0047 \, .	$$

\section{$Z$ boson observables}
%\vglue 0.4cm
%\vglue 0.3cm
%\leftline{\twelveit 1.2. $Z$ boson observables}
%\vglue 0.4cm
%\vglue 1pt
{\it Effective $Z$ boson couplings:}
The predictions for the various
$Z$ widths and asymmetries can conveniently be
calculated in terms of effective coupling constants. They follow
from the set of 1-loop diagrams
without virtual photons.
These weak corrections
can be expressed
in terms of fermion-dependent overall normalizations
$\rho_f$ and effective mixing angles $s_f^2$
in the NC vertices \cite{formfactors}:
\beq
   \left( \sqrt{2}\Gmu\mz \rho_f \right)^{1/2}
\left[ (I_3^f-2Q_fs_f^2)\ganu-I_3^f\ganu\gafi \right]
    =   \left( \sqrt{2}\Gmu\mz \right)^{1/2} \,
  [g_V^f \,\ganu -  g_A^f \,\ganu\gafi]  \, .
\eeq
%The complete expressions for $\rho_f,\kappa_f$ can be found in$^{27}$.
%Up to small terms negligible at the $Z$
%peak, they correspond to those of Bardin et al.$^{28}$.
$\rho_f$ and $s_f^2$ contain  universal
parts     (i.e.\ independent of the fermion species) and
non-universal parts which explicitly depend on the type of the
external fermions.
In their leading terms they are  given by
\beq
\rho_f  =  \frac{1}{1-\drb} + \cdots , \;\;\;
 s_f^2  = s_W^2 + c_W^2\,\drb + \cdots
\eeq
with $\drb$ from Eq.\ (8).

\smallskip
For the $b$ quark, also the non-universal parts have a strong
dependence on $m_t$ resulting from virtual top quarks in the
vertex corrections. The difference between the $d$ and $b$
couplings can be parametrized in the following way
\beq
  \rho_b = \rho_d (1+\tau)^2, \;\;\;\;
  s^2_b = s^2_d (1+\tau)^{-1}
\eeq
with the quantity
$
 \tau = \Delta\tau^{(1)}
      + \Delta\tau^{(2)}
      + \Delta\tau^{(\al_s)}
$
calculated perturbatively, at the present level comprising:
the complete 1-loop order term \cite{vertex}
$
\Delta\tau^{(1)} = -2 x_t + \cdots $;
the electroweak 2-loop contribution of $O(\Gmu^2 m_t^4)$
\cite{barbieri,dhl}
\beq
\Delta\tau^{(2)} = -2\, x_t^2 \, \tau^{(2)} \, ,
\eeq
where
 $\tau^{(2)}$ is a function of $M_H/m_t$
with
 $\tau^{(2)} = 9-\pi^2/3$ for $M_H \ll m_t$;
the QCD corrections to the leading term of $O(\al_s\Gmu m_t^2)$
\cite{jeg}
\beq
\Delta\tau^{(\al_s)} =  2\, x_t \cdot \frac{\al_s}{\pi}
 \cdot \frac{\pi^2}{3} \, .
\eeq

\smallskip
\paragraph{\it Asymmetries and mixing angles:}
The effective mixing angles are of particular interest since
they determine the on-resonance asymmetries via the combinations
   \beq
    A_f = \frac{2g_V^f g_A^f}{(g_V^f)^2+(g_A^f)^2}  \, .
\eeq
Measurements of the \ass hence are measurements of
the ratios
\beq
  g_V^f/g_A^f = 1 - 2 Q_f s_f^2
\eeq
or the effective mixing angles, respectively.

\smallskip
\paragraph{\it $Z$ width and partial widths:}
The total
$Z$ width $\Gamma_Z$ can be calculated
essentially as the sum over the fermionic partial decay widths
(other decay channels are not significant).
Expressed in terms of the effective coupling constants
they read up to 2nd order in the (light) fermion masses:
\bea
\Gamma_f
  & = & \G_0
 \, \left[
     (g_V^f)^2  +
     (g_A^f)^2 \left(1-\frac{6m_f^2}{\mz}\right)
                           \right]
 \cdot   (1+ Q_f^2\, \frac{3\al}{4\pi} )
     \, +\, \Delta\G^f_{QCD} \nn
\eea
with
$$
\G_0 \, =\,
  N_C^f\,\frac{\sqrt{2}\Gmu M_Z^3}{12\pi},
 \;\;\;\; N_C^f = 1
 \mbox{ (leptons)}, \;\; = 3 \mbox{ (quarks)}.
$$
The QCD correction for the light quarks \cite{qcdq}
with $m_q\simeq 0$ is given by
\beq
 \Delta\G^f_{QCD}\, =\, \G_0
  \left[ (g_V^f) ^2+ (g_A^f)^2 \right]
 \cdot    \left[
                  \frac{\al_s}{\pi} +1.41 \left(
  \frac{\al_s}{\pi}\right)^2 -12.8 \left(
  \frac{\al_s}{\pi}\right)^3 \right] \, .
\eeq
For $b$ quarks
the QCD corrections are different due to  finite $b$ mass terms
and to top quark dependent 2-loop diagrams
 for the axial part.
They are calculated up to third order
 in the vector and up to second order
in the axial part \cite{qcdb}.

\smallskip
\paragraph{\it Implications of precision data:}
 In table 1
the \sm predictions for $Z$ pole observables   are
put together. The first error corresponds to
 the variation with $m_t, M_H$
in the range allowed by $M_W$ and $\Dr$ (Fig.\ 1),
the second error is the hadronic
uncertainty from $\al_s=0.123\pm 0.006$ measured
by QCD observables at the $Z$.
 The recent combined LEP results on the $Z$ resonance
parameters \cite{LEP}, under the assumption of lepton universality,
are also shown in table 1, together with $s^2_e$ from
the left-right asymmetry at the SLC \cite{sld}.
The direct information on $m_t$ is not included.

\begin{table}
\bc
\caption{LEP/SLC results and \sm predictions for the $Z$ parameters.}
 \btab{| l | l | r | }
\hline
 observable & LEP/SLC 1994 & \sm range \\
\hline
\hline
$M_Z$ (GeV) & $91.1888\pm0.0044$ &  input \\
\hline
$\Gamma_Z$ (GeV) & $2.4974\pm 0.0038$ & $2.4922 \pm 0.0075\pm 0.0033$ \\
%                 &                  & $2.4933 \pm 0.0064\pm 0.0033$ \\
%\hline
%$\Gamma_{had}$ (GeV) & $1.740\pm 0.008$ &
% $1.736\pm 0.008 \pm 0.007$ \\
%\hline
%$\Gamma_e$ (MeV) & $83.2\pm 0.4$ & $83.7\pm 0.4 $ \\
\hline
$\sigma_0^{had}$ (nb) & $41.49\pm 0.12$ & $41.45\pm0.03\pm0.04$ \\
%                     &                 & $41.45\pm0.01\pm0.04$ \\
\hline
 $\G_{had}/\G_e$ & $20.795\pm 0.040 $ & $20.772\pm 0.028\pm 0.038$ \\
%                &                    & $20.774\pm 0.017\pm 0.038$ \\
%\hline
%$\Gamma_e$ (MeV) & $83.82\pm 0.27$ & $83.7\pm 0.4 $ \\
\hline
$\Gamma_{inv}$ (MeV) & $499.8\pm 3.5$ & $500.8\pm 1.3$ \\
%                    &                & $500.8\pm 0.9$ \\
\hline
$\G_b/\G_{had}$  & $0.2202\pm 0.0020$ & $0.2158\pm 0.0013$ \\
%                &                    & $0.2160\pm 0.0006$ \\
\hline
$\rho_{\ell}$ & $1.0047\pm 0.0022$ & $1.0038\pm 0.0026$ \\
%        &                    & $1.0028\pm 0.0007$ \\
\hline
$s^2_{\ell}$ & $0.2321\pm 0.0004$ & $0.2324\pm 0.0012$ \\
%       &                    & $0.2322\pm 0.0010$ \\
\hline
$s^2_e (A_{LR})$ & $0.2292\pm 0.0010$ & $0.2324\pm 0.0012$   \\
        &  (SLC result)               &                    \\
\hline
\etab
\ec
\end{table}

The $Z$ observables are more constraining to the top mass than $M_W$,
as can be seen from table 1.
Assuming the validity of the \sm a global fit to all electroweak
LEP results yields an indirect determination of
the parameters $m_t,\al_s$ as follows: \cite{LEP}
\beq
    m_t = 173^{+12+18}_{-13-20}\, \gv, \;\;\
    \al_s = 0.126 \pm 0.005 \pm 0.002
\eeq
with $M_H= 300$ GeV for the central value.
The second error is from the variation of $M_H$
between 60 GeV and 1 TeV.
The fit result includes the
uncertainties of the \sm calculations to be discussed in the
next section.

 Including the information on
neutrino scattering and $M_W$
modifies the fit result only marginally \cite{LEP}:
\beq
    m_t = 171^{+11+18}_{-12-21}\, \gv, \;\;\;
    \al_s = 0.126 \pm 0.005 \pm 0.002\, .
\eeq
Incorporating also the SLC result on $A_{LR}$ yields \cite{LEP}
\beq
    m_t = 178^{+11+18}_{-11-19}\, \gv, \;\;\;
    \al_s = 0.125 \pm 0.005 \pm 0.002\, .
\eeq
A simultaneous fit to $m_t$ and $M_H$
 from all low and high energy data but for
constrained $\al_s=0.118\pm 0.007$
 yields a slightly lower range
\cite{jellis}
$ m_t =153 \pm 15$ GeV. For larger values of $\al_s$ the result is
very close to the one in Eq.\ (18) \cite{jellis}.

\section{Status of the Standard Model predictions}
%\vglue 0.4cm
%\vglue 0.3cm
%\leftline{\twelveit 1.4 Status of the Standard Model predictions}
%\vglue 0.4cm
%\vglue 1pt
 For a discussion of the theoretical reliability of the \sm predictions
one has to consider various sources of uncertainties:

The error of the hadronic contribution
to $\al(\mz)$, Eq.\ (4), leads to
$\delta M_W = 13$ MeV in the $W$ mass prediction, and
$\delta\sin^2\theta = 0.0002$ common to all of the mixing
angles, which matches with the future experimental precision.

The uncertainties from the QCD contributions,
 besides the 3 MeV in the
hadronic $Z$ width, can essentially be traced back to
those in the top quark loops for the $\rho$-parameter.
They  can be combined into the following net effects
\cite{kniehl93}
$
 \delta(\dro) \simeq 2\cdot 10^{-4},   \;
 \delta s^2_{\ell} \simeq 1\cdot 10^{-4}
$
for $m_t = 150$ GeV and somewhat larger for heavier top.

The size of unknown higher order contributions can be estimated
by different arrangements of non-leading higher order terms
and investigations of the scheme dependence.
Detailed studies by use of different computer codes,
based on  on-shell and $\ms$ calculations,
for the $Z$ resonance observables
have shown differences around
0.1\%, in particular, $\delta s^2_{\ell} = 1 - 2\cdot 10^{-4}$.
A comprehensive documentation is meanwhile available \cite{ewwgr}.

\section{Conclusions}
 The agreement
of the experimental high and low energy precision
data with the \sm predictions has shown that the \sm works as a fully
fledged quantum field theory.
A great success of the \sm is the directly measured
top mass range which coincides in an
impressive way with
the indirect determination from loop effects in precision data.

The steadily increasing accuracy of the data starts to exhibit
also sensitivity to the Higgs mass \cite{LEP,jellis},
 although still marginally
($M_H< 1$ TeV at 95\% C.L.) for the current situation.

Not understood at present are the deviations
from the theoretical expectation
 observed in the
measurement of $\alr$ and $R_b$, in particular if the CDF top mass
range. Whether they they might be first
hints for non-standard physics makes the future
investigations  particularly exciting.

%\section{References}

\end{document}